\documentclass[aip,apl,graphicx]{revtex4-2}

\usepackage{chemformula} 
\draft 

\begin{document}


\title{Thermal conductivity reduction due to phonon geometrical scattering in nano-engineered epitaxial germanium} 

\author{Jessy Paterson}
\author{Sunanda Mitra}
\author{Yanqing Liu}
\affiliation{Univ. Grenoble-Alpes, CNRS, Grenoble INP, Institut N\'{e}el, 38000 Grenoble, France}
\author{Mustapha Boukhari}
\affiliation{Univ. Grenoble-Alpes, Grenoble INP, CEA, IRIG,	Pheliqs, SiNaPS Lab, Grenoble F-38000, France}
\author{Dhruv Singhal}
\affiliation{Univ. Grenoble-Alpes, CNRS, Grenoble INP, Institut N\'{e}el, 38000 Grenoble, France}
\author{David Lacroix}
\affiliation{Université de Lorraine, CNRS, LEMTA, Nancy F-54000, France}
\author{Emmanuel Hadji}
\author{Andr\'{e} Barski}
\affiliation{Univ. Grenoble-Alpes, Grenoble INP, CEA, IRIG,	Pheliqs, SiNaPS Lab, Grenoble F-38000, France}
\author{Dimitri Tainoff}
\author{Olivier Bourgeois}
\email[Author to whom correspondence should be addressed: ]{olivier.bourgeois@neel.cnrs.fr}
\affiliation{Univ. Grenoble-Alpes, CNRS, Grenoble INP, Institut N\'{e}el, 38000 Grenoble, France}

\date{\today}

\begin{abstract}
Nano-engineering crystalline materials can be used to tailor their thermal properties. By adding new nanoscale phonon scattering centers and controlling their size, one can effectively decrease the phonon mean free path and hence the thermal conductivity of a fully crystalline material. In this letter, we use the 3$\omega$ method in the temperature range of 100-300 K to experimentally report on the more than threefold reduction of the thermal conductivity of an epitaxially-grown crystalline germanium thin film with embedded polydispersed crystalline \ch{Ge3Mn5} nano-inclusions with diameters ranging from 5 to 25~nm. A detailed analysis of the structure of the thin film coupled with Monte Carlo simulations of phonon transport highlight the role of the nano-inclusions volume fraction in the reduction of the phononic contribution to the thermal conductivity, in particular its temperature dependence, leading to a phonon mean free path that is set by geometrical constraints. 
\end{abstract}


\maketitle 

The ever-increasing power density required in today's microelectronic devices is one of the main catalyst behind the continued miniaturization of novel electronic equipment. The implementation of low-dimensional materials in next-generation devices is alas accompanied with great challenges to face –- one of which is the efficient management of heat in size-constrained structures, an essential matter for achieving long-term reliability and preventing rapid failure of devices. From the efficient dissipation of heat away from hot spots for achieving increased working performances in future transistors \cite{Yan2015}, to the localization of heat within a small volume for controlling a material phase in such application as non-volatile phase change memories \cite{LeGallo2020}, a large variety of applications rely upon the thermal properties of their constituting materials. 
	
	Being able to control the thermal properties of nanomaterials is therefore an essential step towards more efficient applications. Engineering a material at scales comparable to the phonon mean free path -- often referred to as "phonon engineering" -- has proven effective in modifying the heat carriers scattering mechanisms and consequently reducing the thermal conductivity\cite{Balandin2005,Kim2006,Biswas2012}.
	Several approaches have been used to modify the ability of a material to conduct heat, around two main axes: on one hand, playing with the geometry of the structure to enhance the scattering efficiency between low-frequency phonons and boundaries such as for nanowires, membranes, and other structures\cite{Li2003,Heron2010,Yu2010,Hopkins2011,Yanagisawa2017,Phys2020,Anufriev2017,Yu2019,Sledzinska2020,Tavakoli2018}. On the other hand, intrinsic modification of materials through doping, alloying or incorporation of nanoparticles/nano-inclusions has been studied both experimentally and theoretically with the intent of increasing phonon scattering between the host and the incorporated materials, \cite{Kim2006,Scotsman2006,Mingo2009,Wang2009a,Mohammed2015,Zhang2015,Dong2019,Nakamura2018,Sakane2019} which is targeted to impede mid-to-high frequency phonons. Using the latter strategy, it is often difficult to obtain a material with scattering centers whose size are a few tens of nanometers, completely crystalline and without affecting the core structure of the host material, \emph{i.e} without creating an alloy. Using nano-voids rather than crystalline nano-inclusions could be easier to elaborate in principle, however such structures are expected to be more sensitive to their environment because of water adsorption in the pores that changes their thermal properties \cite{Isaiev2020}.
	While the effect of alloying and doping has been extensively studied and has proven effective in adding efficient sub-nanometer scattering centers for phonons which reduces the thermal conductivity in thin films \cite{Morelli2002,Wang2009a,Mohammed2015}, the effect of scattering with polydispersed nano-inclusions with size of a few tens of nanometers embedded within a single crystal has been less investigated experimentally, primarily because of the challenges encountered in the growth process. 
	In this letter, we investigate how a distribution of spherical, fully crystalline \ch{Ge3Mn5} nano-inclusions with an average size of 14 nm embedded within a crystalline germanium matrix, which will be henceforth referred to as GeMn thin films, impedes thermal transport as a function of temperature and compare our findings with Monte Carlo (MC) simulations of phonon transport in porous media. 
	
\begin{figure*}
		\centering	
		\includegraphics[width=470pt]{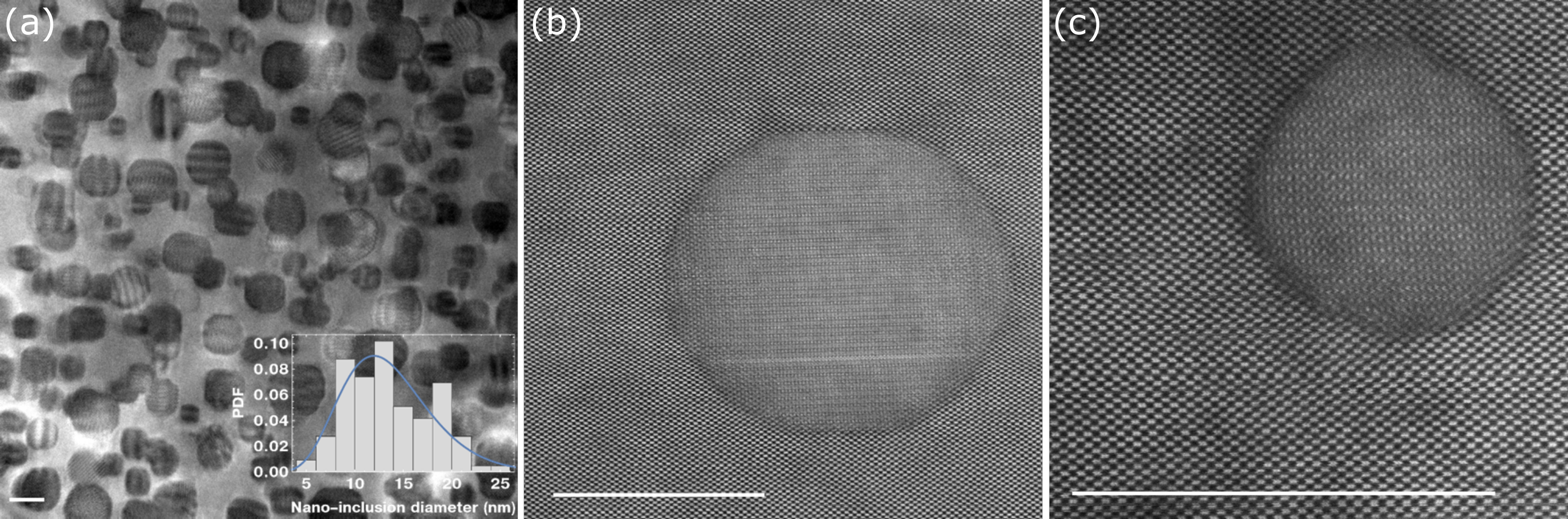}
		\caption{HRSTEM micrographs of the GeMn film with 8\% of Mn. In inset of (a) is the diameter distribution of the nano-inclusions (histogram), fitted using a Gamma distribution (blue line). (b) and (c) show the crystallinity of both the Ge matrix as well as the \ch{Ge3Mn5} nano-inclusions. The white scale bar is 15 nm in all three images.}
		\label{fig:TEM_GeMn}
	\end{figure*}

	We use ultra-high vacuum molecular beam epitaxy (MBE) to grow GeMn thin films. The MBE chamber is equipped with standard effusion cells that permit the films growth via the co-deposition of Ge and Mn. 
	The films were grown on commercially available epi-ready p-type Ge (001) wafers. 
	After the preparation of the surface (Supporting Information, Section 1), a 27 nm thick Ge buffer layer was deposited on the Ge substrate followed by quick annealing at a temperature 314$^\circ$C so as to favor the layer by layer epitaxial growth of GeMn on an atomically flat surface. The GeMn films were epitaxially grown at a low-temperature region of around 92$^\circ$C that is expected to enhance the Mn solubility in Ge and also minimize the phase separation. This low-temperature growth resulted into the formation of auto-organized nano-columns embedded in a crystalline germanium matrix as a result of spinodal decomposition. \cite{Jain2011} After, the annealing was done at a gradual increase of temperature over a certain time period to reach around 400-500$^\circ$C. This step facilitated the transformation of the nano-columns into a more stable state consisting of \ch{Ge3Mn5} spherical nano-inclusions in a Ge matrix. Prior to the annealing step, the GeMn films were capped with an 80 nm germanium layer to avoid the surface roughness possibly occurring due to the migration of manganese upon annealing. In this report, all the samples were grown with a manganese content of nominally 8\% and a GeMn layer of varying thicknesses (360 and 720 nm). During the epitaxial growth process, the quality of the crystalline nature of the films was periodically observed by in-situ reflection-high-electron-energy-diffraction (RHEED) equipped within the MBE chamber (Supporting Information, Section 1). In Figure \ref{fig:TEM_GeMn}, High Resolution Scanning Transmission Electron Microscopy (HRSTEM) analysis shows the crystalline nature of both the germanium matrix and the \ch{Ge3Mn5} nano-inclusions (Supporting Information, Section 2). Besides, an important result of the growth process is the distribution in size of the \ch{Ge3Mn5} nano-inclusions, ranging from 5 to about 25~nm in diameter, which we fitted using a Gamma distribution displayed in the inset of Figure \ref{fig:TEM_GeMn}(a). The size and dispersion of the nano-inclusions are expected to play an important role in modifying the heat carriers mean free path, since both the inter-inclusion distance as well as the inclusion size are smaller than the phonon mean free path of bulk germanium ($\sim$200~nm\cite{Minnich2007,Jean2014}). Size and distribution can be tuned by modifying the germanium to manganese ratio in the germanium matrix during the MBE growth; larger nano-inclusion size and broader distributions are obtained for higher Mn percentage in the film, up to 35~nm for 14\% of Mn, keeping the high crystalline quality.
	
	We measure the thermal conductivity of the GeMn thin films in the temperature range of 100-300~K using the 3$\omega$ method\cite{Cahill1990}, for one particular nano-inclusions distribution. The 3$\omega$ method consists in measuring the thermal impedance created by a specimen of interest, when submitted to a periodic heat flux. The periodic heat flux is created by applying an alternative current of frequency 1$\omega$ across a metallic line microfabricated on top of the specimen. Joule heating occurring at 2$\omega$ induces a temperature oscillation at the same frequency, which, coupled with the metallic line's temperature coefficient of resistance, leads to a periodic change of the metallic line resistance at 2$\omega$. Consequently, a voltage at frequency 3$\omega$ that is proportional to the temperature oscillation $\Delta T_{2\omega}$ develops across the metallic line. It is from this temperature oscillation that we infer thermal properties of the system beneath the metallic line. Varying the oscillation frequency allows to probe in-depth the system of interest, and fitting the frequency-dependence of the measured temperature oscillation to an adequate thermal model allows to extract the thermal properties of the system studied. The metallic line consists of a nominally 100 nm thick, $\sim$5.5 $\mu$m wide platinum thin film deposited using magnetron sputtering -- width that have been measured using scanning electron microscopy. Since the GeMn samples are electrically conductive, they have been capped using a $60$ nm thick \ch{Al2O3} layer deposited by Atomic Layer Deposition prior to the deposition of the metallic line to ensure electrical insulation of the thermometers.
	
\begin{figure*}
		\centering	
		\includegraphics[width=470pt]{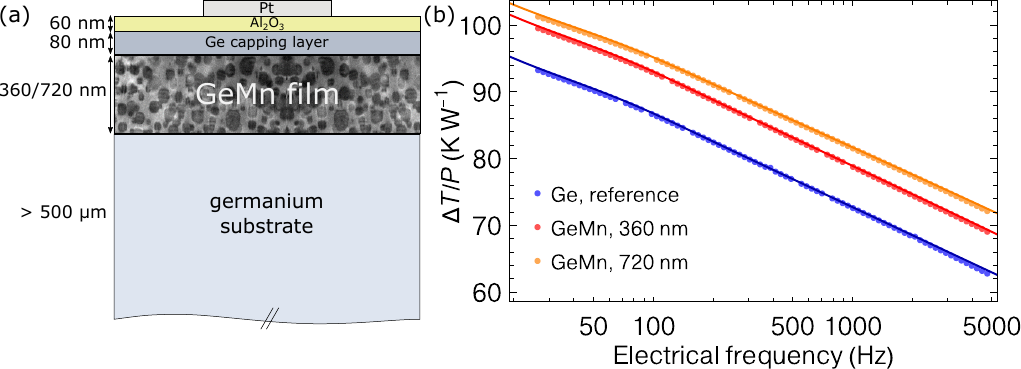}
		\caption{(a) Schematic of the stacking used to conduct 3$\omega$ measurements on the GeMn samples, with the Pt heater deposited on top of the sample. The reference germanium sample corresponds to a germanium substrate with a 60 nm thick dielectric \ch{Al2O3} layer. (b) Absolute value of the temperature oscillation normalized to heating power measured as a function of frequency using the 3$\omega$ method for the GeMn and reference samples, at 300~K. Coloured filled circles correspond to experimental data while full lines correspond to the best fit of the data to Eq.~\ref{eqn:deltaT}.}
		\label{fig:DeltaT}
	\end{figure*}

	The complexity of the measurement arises from the relatively low thermal conductivity contrast between the GeMn film and the germanium substrate, leading to both a small thermal signal from the film and a difficulty to separate its contribution from other sources of thermal signals, such as thermal boundary resistances (TBRs) present in the system, and the dominant thermal contribution of the germanium substrate. To overcome these difficulties, we have used a combination of sensitive measurements and advanced data modelling and uncertainty analysis. The 3$\omega$ measurements have been performed using a differential bridge as described in our previous work\cite{Paterson2020} in order to reach a sufficient signal-to-noise ratio. Then, as the commonly used simple one-dimensional thermal model is not relevant in this case for inferring the thermal conductivity of the GeMn film -- because of its relatively high thermal conductivity -- we use a thermal model based on Borca-Tasciuc \emph{et al.} work \cite{Borca-Tasciuc2001} to fit the frequency-dependence of the temperature oscillation for our multilayer system, which is depicted in Figure \ref{fig:DeltaT}(a). We write the temperature oscillation as follows:
	\begin{subequations}\label{eqn:deltaT}
		\begin{equation}\label{eqn:deltaTa}
			\Delta T=\frac{-P}{\pi l k_{y_1}}\int_0^{\infty}\frac{1}{A_1B_1}\frac{\sin^2(b\lambda)}{\left(b\lambda\right)^2}d\lambda+\frac{P}{2bl}\times R_{\left(\text{TBRs +\ch{Al2O3}}\right)}
		\end{equation}
		where 
		\begin{equation}
			A_{i-1}=\frac{A_i\frac{k_{y_i}B_i}{k_{y_{i-1}}B_{y_{i-1}}}-\tanh (B_{i-1}d_{i-1})}{1-A_i\frac{k_{y_i}B_i}{k_{y_{i-1}}B_{y_{i-1}}}\tanh (B_{i-1}d_{i-1})},\quad i=2,..,n
		\end{equation}
		with
		\begin{equation}
			B_i=\left(k_{xy_i}\lambda^2+i 2\omega\frac{ \rho_i C_{p_i}}{k_{yi}}\right)^{1/2},\quad k_{xy_i}=\frac{k_{x_i}}{k_{y_i}}
		\end{equation}
	\end{subequations}
	$P$, $l$, $k_{x_i}$, $k_{y_i}$, $b$, $d_i$, $\rho_i$, and $C_{p_i}$ refer to the electrical power, heater length, in-plane thermal conductivity of layer $i$, cross-plane thermal conductivity of layer $i$, heater half-width, thickness of layer $i$, density of layer $i$ and heat capacity of layer $i$, respectively. $n$ refers to the total number of layers. The thermal contribution of the 60 nm \ch{Al2O3} film and of the Pt/\ch{Al2O3} + \ch{Al2O3}/Ge TBRs are grouped as one single thermal resistance labeled as $R_{\left(\text{TBRs +Al$_2$O$_3$}\right)}$, while the three other layers (germanium substrate, GeMn layer and the 80 nm germanium capping layer) are accounted for in the left hand side term of Eq.~\ref{eqn:deltaTa} and therefore $n=3$ in Eq.~\ref{eqn:deltaT}.
	The thermal contribution of the 60 nm \ch{Al2O3} film and of the Pt/\ch{Al2O3} + \ch{Al2O3}/Ge interfaces is consequent and has been measured separately, using a 60 nm \ch{Al2O3} film deposited on germanium that has been deposited during the same run as that deposited on the GeMn samples. This additional sample will be referred to as the reference sample in the following.  
 
	The specific heat capacity of germanium has been measured as a function of temperature using Quantum Design's PPMS\textsuperscript \textregistered (Supporting Information, Section 4) while that of manganese is taken from literature \cite{Desai1987}. We assume all layers to have no thermal anisotropy and set $k_{xy}=1$. The remaining unknowns in the thermal model are the thermal conductivity of the germanium substrate, GeMn sample and germanium capping layer. The thermal conductivity of the germanium substrate is determined in a straightforward manner from the slope of the temperature oscillation as a function of frequency, and we take $k_{\text{Ge, 80 nm}}\sim 35~$W m$^{-1}$ K$^{-1}$ from literature\cite{Alvarez-Quintana2011,Wang2011d} for the 80 nm thick germanium capping layer. This value of thermal conductivity is reduced as compared to the bulk value of 53 W m$^{-1}$ K$^{-1}$ due to phonon boundary scattering. We use this literature value rather than measuring it because the thermal resistance from an 80~nm thin film with a thermal conductivity of 35 W m$^{-1}$ K$^{-1}$ would lead to a signal that is below our measurement resolution in this configuration (supported film rather than suspended one). The uncertainty for this thermal conductivity value is taken into account in the uncertainty calculation (Supporting Information, Section 3).
	
	In order to study a possible effect of the thickness on the thermal conductivity of the GeMn films, we have performed measurements on films with two different thicknesses, 360 and 720 nm. The measurement on the thickest film is expected to be less affected by boundary scattering that can reduce the phonon thermal conductivity as well, as the phonon mean free path in germanium is expected to be of the order of 200~nm\cite{Minnich2007,Jean2014}. In Figure \ref{fig:DeltaT}(b), the temperature oscillation normalized to heating power measured at 300~K using the 3$\omega$ method is plotted as a function of electrical frequency for the two GeMn samples and the reference sample, and fitted using Eq.~\ref{eqn:deltaT}. The fitting is very satisfying over the entire frequency range, and we obtain the same value of the thermal conductivity of the germanium substrate within 4\% of uncertainty for the three samples, which underlines the repeatability of the experiment. We note that even though the thermal weight of the GeMn film is relatively low as compared to other sources of thermal resistance, we observe a clear, confidently distinguishable, increase of the thermal resistance between the reference sample and the GeMn films as well as between the two GeMn samples with different thicknesses, which translates as an upward shift of the temperature oscillation in Figure \ref{fig:DeltaT}(b).
	
\begin{figure}
		\centering
		\includegraphics[width=240pt]{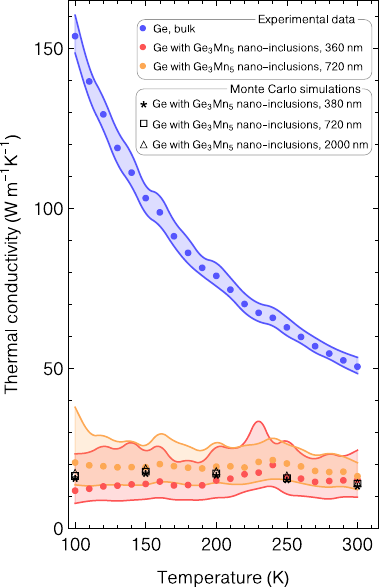}
		\caption{Thermal conductivity as a function of temperature of the GeMn thin films consisting of crystalline \ch{Ge3Mn5} spherical nano-inclusions embedded within a germanium matrix, measured using the 3$\omega$ method. The thermal conductivity of the germanium substrate onto which the GeMn layers are grown is displayed (in blue) for comparison : a reduction in thermal conductivity by a factor of ten is observed at 100~K for the 360 nm thick film. Tinted bands represent 95\% Confidence Intervals of the reported thermal conductivity. Black stars, squares and triangle represent the calculated thermal conductivity of the GeMn films using a Monte Carlo method, for three different thicknesses.}
		\label{fig:kGeMn}
	\end{figure}

	As the uncertainty of some of the parameters in the thermal model is quite large, we cannot use a standard method for the propagation of uncertainty and instead we use a Monte Carlo method to estimate the uncertainty of the thermal conductivity of the GeMn films\cite{Saltelli2007,Chen2018a,Paterson2020}. Tinted bands around the thermal conductivity of the GeMn and reference samples in Figure \ref{fig:kGeMn} represent 95\% confidence intervals obtained from fitting experimental data to Eq.~\ref{eqn:deltaT} and applying an uncertainty analysis based on the Monte Carlo method (Supporting Information, Section 3).

	In addition to the reduction of the thermal conductivity of the GeMn films by a factor of more than three at room temperature compared to bulk germanium -- from 51 to $\sim$ 16 W m$^{-1}$K$^{-1}$ -- the most important result of this work is the measured temperature dependence of the GeMn thermal conductivity, which is almost flat from 100 to 300~K. This suggests that the addition of crystalline spherical nano-inclusions leads to an additional phonon scattering mechanism, which is dominant over other scattering mechanisms that are present in germanium in this temperature range, such as scattering with isotopes, electrons, impurities and other phonons \cite{Asen-Palmer1997}-- and leads to a constant mean free path. In an attempt to go further in the understanding of the phonon scattering mechanisms governing the thermal conductivity in the GeMn films, we turn to a numerical approach of phonon transport using a MC method. 
	
\begin{figure*}
		\centering
		\includegraphics[width=470pt]{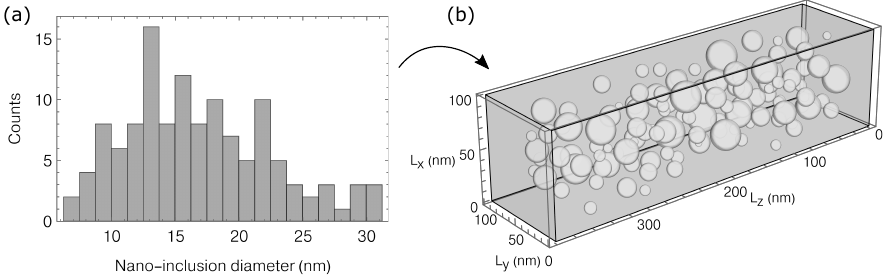}
		\caption{Nano-inclusion size distribution used for calculating the GeMn thin film thermal conductivity using a Monte Carlo method. (a) Example of one of the four sampled nano-inclusion size distributions, obtained from a Gamma distribution. (b) Example of an initial configuration of the thin film geometry consisting of a host germanium matrix (gray) with polydispersed spherical nano-inclusions (light gray).}
		\label{fig:MC_schema}
	\end{figure*}

	Numerical modeling of heat transfer in structures with complex geometries, such as germanium thin films with polydisperse nanometric inclusions presented in this study, is a challenge. On one hand, the tools developed for continuous media cannot be used in these systems where the mean free path of the energy carriers, the phonons, is much lower than in the bulk material. On the other hand, an atomistic approach such as molecular dynamics is difficult to implement to describe these objects. On one side, the films considered have a size that would require computational resources far exceeding those commonly available in computing centers. On the other side, as we have shown in a previous study \cite{Mangold2020}, there are no "simple" interatomic potentials characterizing the interactions between Ge and Mn atoms. Only machine learning approaches are able to model these interactions, which makes the modeling even more complex. In this context, numerical tools based on the statistical solution of the Boltzmann transport equation for phonons in the relaxation time approximation are an alternative. The MC method used in this work has been validated multiple times when numerically modeling phonon transport in nanostructured semiconductors \cite{Lacroix2005a,Jean2014a}. However, the detailed modeling of \ch{Ge3Mn5} inclusions remains a challenge, in particular because there is no data on the intrinsic collisional lifetimes (Normal and Umklapp processes). Therefore, the proposed calculations make the assumption that the inclusions are similar to "opaque" structures on the surface of which phonons are diffusely reflected. There is no other assumption, the modeled films have thicknesses close to those obtained experimentally (380~nm, 720~nm and 2000~nm). The size distribution of the inclusions is randomly drawn according to a Gamma distribution constructed using the parameters identified experimentally ($\alpha$ = 8.92, $\beta$ = 0.93) which corresponds to a 12.7\% inclusion volume fraction in the case where the samples are grown with 8\% Mn. An illustration of this type of structure is given in Figure \ref{fig:MC_schema}(b). The thin film is discretized by 100~nm wide cells. The phonon reflections on the lateral faces of the cells are specular which guarantees the periodicity of the structure and thus the calculation of the transport properties is done in the cross-plane direction ($L_z$ in Figure \ref{fig:MC_schema}(b)) which is the quantity that is measured experimentally. The phonon flow is induced by a temperature difference of 2~K, around the average temperature, which is imposed between the extreme cells of the system. These are treated as phonon blackbodies. Finally, each calculation is averaged over four initial configurations of nano-inclusion distributions to reduce the statistical bias.
	The MC simulations have been performed in the temperature range of 50 to 400~K, covering the experimentally explored domain with a calculation point every 50~K. Three thicknesses were tested, 380, 720 and 2000~nm. The thermal conductivities deduced from these MC simulations are reported as black shapes in Figure~\ref{fig:kGeMn}. We observe a very small variation of the thermal conductivity over the considered range and the experimentally observed "plateau" is well recovered. Moreover, the impact of the film thickness is also modest. There is a slight increase in the thermal conductivity as the film becomes thicker, but this remains much less marked than for a thin film without inclusion. As an example in the case of Ge, the calculated thermal conductivity of the 380, 720 and 2000~nm pristine films, at 300K, is respectively:  38, 43 and 52 W m$^{-1}$ K$^{-1}$. We can therefore deduce that the transport is mainly dominated by the collisions of phonons with the inclusions, leading to a phonon mean free path that is set by geometrical effects.
	\begin{figure}
		\centering
		\includegraphics[width=240pt]{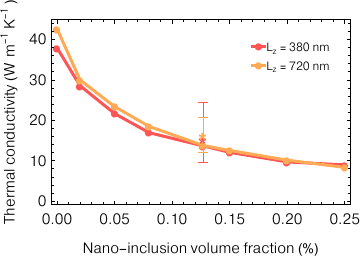}
		\caption{Impact of the nano-inclusion volume fraction on the thermal conductivity of germanium thin films with polydispersed nano-inclusions, calculated using the MC method, for two thicknesses of the film. The two stars represent experimental measurements performed on the GeMn films at 300~K for a measured volume fraction of 12.7\%.}
		\label{fig:k_vs_poros}
	\end{figure}
	The relevance of numerical simulation in studying phonon transport lies in its ability to provide insights into the behavior of such structures. It is thus possible, at low cost, to evaluate the properties of thin films containing a more or less important fraction of Mn and to observe the impact of the enrichment on the thermal conductivity of the resulting structures. In Figure \ref{fig:k_vs_poros}, we show the evolution of the thermal conductivity of thin films (380 and 720~nm) at 300~K with inclusions whose size distribution follows the Gamma law discussed above but whose \ch{Ge3Mn5}  inclusion volume fraction varies continuously from 0 to 25\%. As the inclusion volume fraction increases, the thermal conductivity logically decreases, however we note that this trend is not linear. Doubling the amount of inclusions leads to a reduction of the thermal conductivity but in a smaller proportion than the reduction observed when going from 0 to 12.7\%. However, it is important to keep in mind that the nano-inclusion size distribution is unchanged. 
	
	In summary, a model material consisting of a germanium matrix with spherical \ch{Ge3Mn5} nano-inclusions has been grown using MBE and fully characterized using TEM, showing the crystallinity of both the host matrix and nano-inclusions. Sensitive 3$\omega$ measurements and data processing have enabled the observation of the reduction by a factor of ten at 100K, and three at 300K, of the thermal conductivity of the fully crystalline nanostructured film compared to its non-structured counterpart, and more importantly reveal a plateau in the temperature dependence of the thermal conductivity. This suggests that the crystalline nano-inclusions within the germanium matrix set the phonon mean free path to a constant value. Numerical phonon transport simulations of thermal transport in porous germanium using a MC method reproduce very well this behavior, suggesting that while the nano-inclusions are crystalline, they completely diffuse incoming phonons at their interface -- a promising result for potential phonon-glass electron-crystal semiconducting materials. Besides, a noteworthy finding of this study on fully crystalline materials is that the nature of inclusions (crystalline, amorphous, or void) appears to be less significant than their volumetric concentration and size. Indeed, our Monte Carlo simulations effectively support the observed reduction in thermal conductivity and temperature behavior, where nano-inclusions are treated as voids rather than fully crystalline inclusions.
	The different synthesis campaigns carried out on the GeMn thin films have shown that the initial mass fraction of Mn has a significant impact on the size and volume fraction of the nano-inclusions within the germanium matrix, which could further reduce its thermal conductivity \cite{Kim2006a} -- a more detailed study on this subject is in progress.

 See the supplementary material for details. The supplementary materials include: (1) RHEED pattern of GeMn thin films during growth, (2) TEM analysis of the GeMn films, (3) detailed procedure for uncertainty calculation of 3$\omega$ measurements, (4) specific heat capacity of germanium.

\begin{acknowledgments}
The authors thank the technical support provided by Institut N\'eel, the Pole Capteur, especially E. Andr\'e and G. Moiroux for the platinum deposition, Nanofab for clean room processes. This work is supported by the Agence Nationale de la Recherche by the MESOPHON project grant number ANR-15-CE30-0019 and the Laboratoire d'excellence LANEF in Grenoble (ANR-10-LABX-51-01).
\end{acknowledgments}
\section*{Author declarations}
\subsection*{Conflict of Interest}
The authors have no conflict to disclose.
\subsection*{Author Contributions}
\textbf{Jessy Paterson}: Conceptualization (equal); Methodology (equal); Data Curation (lead); Formal Analysis (equal); Visualization (lead); Writing/Original Draft Preparation (lead).
\textbf{Sunanda Mitra}: Writing/Review \& Editing (equal).
\textbf{Yanqing Liu}: Writing/Review \& Editing (equal).
\textbf{Mustapha Boukhari}: Writing/Review \& Editing (equal).
\textbf{Dhruv Singhal}: Writing/Review \& Editing (equal).
\textbf{David Lacroix}: Conceptualization (equal); Funding Acquisition (equal); Methodology (equal); Formal Analysis (equal); Software (lead); Writing/Review \& Editing (equal).
\textbf{Emmanuel Hadji}: Writing/Review \& Editing (equal).
\textbf{André Barski}: Writing/Review \& Editing (equal).
\textbf{Dimitri Tainoff}: Conceptualization (equal); Writing/Review \& Editing (equal).
\textbf{Olivier Bourgeois}: Conceptualization (equal); Funding Acquisition (equal); Methodology (equal); Formal Analysis (equal); Writing/Review \& Editing (equal); Supervision (lead).
\section*{Data Availability}
The data that support the findings of this study are available from the corresponding author upon reasonable request.
\bibliography{Paper_GeMn_sub_APL.bbl}

\end{document}